\documentclass[aps,pre,twocolumn,superscriptaddress,showpacs]{revtex4-1}
\usepackage{graphicx}
\usepackage{amsmath}
\usepackage{natbib}
\usepackage[usenames]{color}
\usepackage{soul}

\usepackage{amsfonts}
\usepackage{dcolumn}        
\usepackage{amssymb}
\usepackage{bm,amssymb}

\usepackage{bm,fancybox}                	     

\begin{document}

\newcommand{\bea}{\begin{eqnarray}}
\newcommand{\eea}{  \end{eqnarray}}
\newcommand{\bit}{\begin{itemize}}
\newcommand{\eit}{  \end{itemize}}

\newcommand{\be}{\begin{equation}}
\newcommand{\ee}{\end{equation}}
\newcommand{\ra}{\rangle}
\newcommand{\la}{\langle}
\newcommand{\U}{\widetilde{U}}

\def\bra#1{{\langle#1|}}
\def\ket#1{{|#1\rangle}}
\def\bracket#1#2{{\langle#1|#2\rangle}}
\def\inner#1#2{{\langle#1|#2\rangle}}
\def\expect#1{{\langle#1\rangle}}
\def\e{{\rm e}}
\def\proj{{\hat{\cal P}}}
\def\tr{{\rm Tr}}
\def\H{{\hat H}}
\def\Hdag{{\hat H}^\dagger}
\def\Lop{{\cal L}}
\def\Ehat{{\hat E}}
\def\Edag{{\hat E}^\dagger}
\def\Shat{\hat{S}}
\def\Sdag{{\hat S}^\dagger}
\def\Ahat{{\hat A}}
\def\Adag{{\hat A}^\dagger}
\def\U{{\hat U}}
\def\Udag{{\hat U}^\dagger}
\def\Zhat{{\hat Z}}
\def\Phat{{\hat P}}
\def\Op{{\hat O}}
\def\id{{\hat I}}
\def\x{{\hat x}}
\def\P{{\hat P}}
\def\Px{\proj_x}
\def\Pr{\proj_{R}}
\def\Pl{\proj_{L}}


\title{The role of short periodic orbits in quantum maps with continuous openings}

\author{Carlos A. Prado}
\affiliation{Comisi\'on Nacional de Energ\'{\i}a At\'omica, 
Departamento de F\'{\i}sica. 
Av.~del Libertador 8250, 1429 Buenos Aires, Argentina}
\affiliation{Departamento de F\'{\i}sica, FCEyN, Universidad de Buenos Aires, 
Ciudad Universitaria, Buenos Aires 1428, Argentina.}

\author{Gabriel G. Carlo}
\email[E--mail address: ]{carlo@tandar.cnea.gov.ar}
\affiliation{Comisi\'on Nacional de Energ\'{\i}a At\'omica, CONICET, 
Departamento de F\'{\i}sica. 
Av.~del Libertador 8250, 1429 Buenos Aires, Argentina}

 \author{R. M. Benito}
\affiliation{Grupo de Sistemas Complejos
 and Departamento de F\'{\i}sica,
 Escuela T\'ecnica Superior de Ingenieros
 Agr\'onomos, Universidad Polit\'ecnica de Madrid,
 28040 Madrid, Spain}

\author{F. Borondo}
\affiliation{Departamento de Qu\'{\i}mica, and
 Instituto de Ciencias Matem\'aticas (ICMAT),
 Universidad Aut\'onoma de Madrid,
 Cantoblanco, 28049--Madrid, Spain}
\date{\today}

\begin{abstract}

We apply a recently developed semiclassical theory of short periodic orbits 
to the continuously open quantum tribaker map. In this paradigmatic system the trajectories are partially 
bounced back according to continuous reflectivity functions. This is relevant in many 
situations that include optical microresonators and more complicated 
boundary conditions. In a perturbative regime, the shortest periodic orbits belonging to the classical repeller 
of the open map -- a cantor set given by a region of exactly zero reflectivity -- prove to be 
extremely robust in supporting a set of long-lived resonances of the continuously open quantum maps. 
Moreover, for step like functions a significant reduction in the number needed is obtained, similarly to 
the completely open situation. This happens despite a strong change in the spectral properties when 
compared to the discontinuous reflectivity case.
\end{abstract}
\pacs{05.45.Mt, 03.65.Sq}
\maketitle

\section{Introduction}
 \label{sec:intro}

 In many experimental situations like in the case of optical cavities \cite{RecentExp,WiersigR,Microlasers}, 
the knowledge of properties of an open system becomes crucial. This is also a very interesting 
theoretical problem, even from a pure mathematical point of view \cite{Dyatlov}. 
At the classical level, these situations are usually modeled by eliminating all the 
trajectories that arrive at a given region of phase space (the opening) giving rise 
to a fractal invariant set, the repeller. The quantum analogues of these systems are 
characterized by a set of resonances and the number of long-lived states scales 
with the Planck constant as $\hbar^{-d/2}$, where $d+1$ is the fractal dimension
of the classical repeller. This is the so called fractal Weyl law 
\cite{conjecture,Nonnenmacher,Hamiltonians,qmaps}.
However, the reflection mechanisms at the boundaries are usually more complicated than this complete opening. 
The first step in order to understand these mechanisms is considering a constant reflectivity $R$, meaning that the 
classical trajectories arriving to the opening are partially reflected. 
In these cases we have a multifractal behaviour \cite{Ott} and the usual fractal Weyl law needs to be 
non trivially modified. This has been done for the case of maps \cite{Altmann}, 
which are very suitable models for more complicated systems.

There is another point of view to study this problem which is based on the semiclassical theory of 
short periodic orbits (POs) for open quantum maps \cite{art0}. The shortest POs contained in the 
classical repeller are used to construct a basis set of scar functions which expands the quantum repeller 
and is suitable to express the quantum non-unitary operators \cite{art1,art2,art3}. 
Recently, we have extended the short POs theory to partially open quantum maps where 
a fraction of the quantum probability is reflected \cite{art4}.

The next step for a better understanding of the properties derived from more realistic reflection mechanisms 
at the boundary \cite{RecentExp} is to consider a reflectivity function. In this paper we apply our 
semiclassical theory to a continuously open tribaker map. Although we will drop the word ``partially'' 
from here on in order to simplify the notation, this is obviously a particular case of a partially open map. 
We take into account a step function of the Fermi-Dirac kind in order to smooth the boundaries of the opening, 
and a sinusoidal function which provides with a more generic profile. 
We have found strong changes in the spectral behaviour with respect to 
the discontinuous openings. Despite this, the shortest POs belonging to the classical repeller (corresponding 
to the fully open scenario) explain the main properties of these maps, in a perturbative regime. 
Moreover, in the step function case there is still a significant reduction 
in the number of POs needed for the semiclassical calculations. 

 This paper is organized as follows: 
In Sec. \ref{sec:COpenmaps} we make a brief description of our semiclassical approach and also 
define the system used, i.e. the classical and quantum continuously open tribaker map. 
In Sec. \ref{sec:results} we apply the short POs theory and discuss the results. 
Our conclusions are presented in Sec. \ref{sec:conclusions}

\section{Continuously open maps}
 \label{sec:COpenmaps}

Classical and quantum chaos have benefited from the study of maps, which capture 
all the essential properties of more complicated dynamical systems \cite{Ozorio 1994,Hannay 1980,Espositi 2005}. 
Open maps are transformations of the 2-torus where trajectories disappear when they 
reach an open region in the bidimensional phase-space. The intersection of the forward and backwards trapped sets 
(trajectories that do not escape either in the past or in the future) form the repeller, an 
invariant of fractal dimension. Partially open maps are those in which the opening does not absorb 
all the trajectories that arrive at it, but reflects back a certain amount. Previously \cite{art4}, we 
have investigated a constant function given by a reflectivity $R \in [0:1]$. Here, we consider 
two different functions of the phase space, $F_R(q,p)$, where now $R$ is a parameter that performs 
the transition between the minimum amount of reflection $R=0$ (which is not the completely open case) and 
the closed system ($R=1$). This is intended to capture the general properties of optical 
microcavities and more complicated boundary conditions.

Multifractality manifests itself through a measure that now is not uniformly distributed on the repeller. 
In each phase space region $X_i$, this measure depends on 
the average intensity $I_t$ when $t \rightarrow \infty$ of a number $N_{\rm ic}$ of 
random initial conditions taken inside $X_i$. The initial intensity is $I_0=1$ for each 
trajectory, and changes to $I_{t+1}=F_R(q,p) I_t$ each time it hits the opening. 
The finite time measure for $X_i$ is given by $\mu_{t,i}^{b}=\langle I_{t,i} \rangle/\sum_i 
\langle I_{t,i} \rangle$ where the average is over the initial conditions in the given phase space region. 
If we consider this to be the analogue of the backwards trapped set of open maps, we can 
evolve backwards and obtain $\mu_{t,i}^{f}$ the analogue of the forward trapped set. Their intersection 
gives what we call the {\em continuous repeller} $\mu_{t,i}$.
 
To quantize a map we impose boundary conditions for
both the position and momentum representations by taking 
$\bracket{q+1}{\psi}\:=\:e^{i 2 \pi \chi_q}\bracket{q}{\psi}$, and
$\bracket{p+1}{\psi}\:=\:e^{i 2 \pi \chi_p}\bracket{p}{\psi}$, with $\chi_q$, $\chi_p \in
[0,1)$. In a Hilbert space of finite dimension $N=(2 \pi \hbar)^{-1}$, the
semiclassical limit corresponds to $N \rightarrow \infty$, and the propagator is given by a
$N\times N$ matrix. Position and momentum eigenstates are given by
$\ket{q_j}\:=\:\ket{(j+\chi_q)/N}$ and $\ket{p_j}\:=\:\ket{(j+\chi_p)/N}$ with
$j\in\{0,\ldots, N-1\}$. A discrete Fourier transform gives 
$\bracket{p_k}{q_j}\:=\: \frac{1}{\sqrt{N}} e^{-2i\pi(j+\chi_q)(k+\chi_p)/N} \: \equiv \:
(G^{\chi_q, \chi_p}_N)$. The opening (we take a strip parallel to the $p$ axis) is quantized 
as a projection operator $P$ on its complement, so the open map is of the general form 
$\widetilde{U}=PUP$, where $U$ is the propagator for the closed one. 
Here we take an opening function so the projector becomes $\sqrt{F_R} \times \openone$, 
where the identity has the dimension of the escape region. 
This map has $N$ right eigenvectors $|\Psi^R_j\ra$ and $N$ left
ones $\la \Psi_j^L|$, which are mutually orthogonal $\la
\Psi_j^L|\Psi^R_k\ra=\delta_{jk}$, and that are associated to resonances $z_j$. Our normalization is such 
that $\la\Psi_j^R|\Psi^R_j\ra=\la \Psi_j^L|\Psi^L_j\ra$.

\subsection{Semiclassical theory}

We have recently developed a semiclassical theory \cite{art4} that can be directly applied 
to obtain the resonances of continuously open maps by means of their shortest POs. 
We now give a brief description of the main details. Let $\gamma$ be a PO of fundamental 
period $L$ that belongs to a continuously open map. 
We can define coherent states $|q_j,p_j\rangle$ associated to each point of the orbit. 
We then construct a linear combination 
with them: \begin{equation} |\phi_\gamma^m\rangle=\frac{1}{\sqrt{L}}\sum_{j=0}^{L-1}
\exp\{-2\pi i(j A^m_\gamma-N\theta_j)\}|q_j,p_j\rangle,\end{equation} where 
$m\in\{0,\ldots,L-1\}$ and $\theta_j=\sum_{l=0}^j S_l$. In this expression $S_l$ is the action acquired 
by the $l$th coherent state in one step of the map. The total action is $\theta_L\equiv S_\gamma$
and $A^m_\gamma=(NS_\gamma+m)/L$. Finally, the right and left scar functions for the PO 
are defined through the propagation of these linear combinations under the continuously open map 
$\widetilde{U}$ (up to approximately the system's Ehrenfest time $\tau$). 
\begin{equation}\label{prop}
|\psi^R_{\gamma,m}\rangle=\frac{1}{\mathcal{N}_\gamma^R}\sum_{t=0}^{\tau}
\widetilde{U}^te^{-2\pi iA^m_\gamma t}\cos\left(\frac{\pi
t}{2\tau}\right)|\phi_\gamma^m\rangle,\end{equation} and \begin{equation}
\langle\psi^L_{\gamma,m}|=\frac{1}{\mathcal{N}_\gamma^L}\sum_{t=0}^{\tau}
\langle\phi_\gamma^m|\widetilde{U}^te^{-2\pi iA^m_\gamma t}\cos\left(\frac{\pi
t}{2\tau}\right).\end{equation} Normalization ($\mathcal{N}_{\gamma}^{R,L}$) 
is chosen in such a way that $\langle \psi_{\gamma,m}^R|\psi^R_{\gamma,m}\rangle=
\langle \psi_{\gamma,m}^L|\psi^L_{\gamma,m}\rangle$ and
$\langle \psi_{\gamma,m}^L|\psi^R_{\gamma,m}\rangle=1$.
We then select a number of POs, $N^{POs}$, from the whole set up to a period $L$, 
in order to cover the continuous repeller. In this work we use all POs up to period $L$ that are 
inside the repeller (of the fully open case). We also consider a few of them that are outside, 
$N^{outPOs}_{max}$, having the greatest values of $\mu$, and 
optimized to provide with the most uniform covering possible of the continuous repeller. 
We form a basis set in which we express the continuously open evolution operators 
 $\langle\psi_{\alpha,i}^L|\widetilde{U}|\psi_{\beta,j}^R\rangle$. 
Taking into account $\langle \psi_n^L|\psi_m^R\rangle\neq\delta_{nm}$, 
we solve a generalized eigenvalue problem to obtain the eigenstates. 
The long-lived resonances \cite{art1} are the linear combination of the scar functions of the basis using 
the eigenvector's coefficients. 

\subsection{The tribaker map}

All calculations are preformed on the tribaker map 
\begin{equation}
\mathcal B(q,p)=\left\{
  \begin{array}{lc}
  (3q,p/3) & \mbox{if } 0\leq q<1/3 \\
  (3q-1,(p+1)/3) & \mbox{if } 1/3\leq q<2/3\\
  (3q-2,(p+2)/3) & \mbox{if } 2/3\leq q<1\\
  \end{array}\right.
\label{classicaltribaker}
\end{equation}
This is an area-preserving, uniformly hyperbolic, piecewise-linear and invertible map with
Lyapunov exponent $\lambda=\ln{3}$. The opening region is the domain $1/3<q<2/3$ of the 
reflectivity function $F_R$. We use two kinds of functions, in the first place we consider 
$F_R(q,p)=(1-R)/(1+\exp(-A(q-B)))+R$ for $q>1/2$ and $F_R(q,p)=(1-R)/(1+\exp(-A((1-q)-B)))+R$ for $q<1/2$, 
which is a step function of the Fermi-Dirac kind. We take $A=120$ and $B=0.63$ 
which corresponds to approximately a value $1$ at $q=1/3$ and $q=2/3$ and a bottom at 
$R$ in the middle of the opening region. This function aims to smoothing the transition 
between the closed and the open regions of the phase space, but retaining at the same time 
some of the properties of the discontinuous case, like the flat bottom and the quick drop in reflectivity. 
The other function is given by $F_R(q,p)=((1-R) \cos(6 \pi q)+(1+R))/2$, which is essentially a 
sinusoid that matches a value $1$ at the boundaries of the opening and a minimum given by $R$. 
This function is intended to capture the main properties of optical setups like in microlasers 
experiments.

The quantum version uses the discrete Fourier transform $G^{1/2,1/2}_N$ with 
antiperiodic boundary conditions ($\chi_q=\chi_p=1/2$) to preserve time reversal 
and parity. In position representation the tribaker map is given by \cite{Saraceno1,Saraceno2}
\begin{equation}\label{quantumbaker}
 U^{\mathcal{B}}=G_{N}^{-1}G_{N/3}, 
\end{equation}
with
\begin{equation}\label{quantumbaker2}
 G_{N/3}=\left(\begin{array}{ccc}
  G_{N/3} & 0 & 0\\
  0 & G_{N/3} & 0\\
  0 & 0 & G_{N/3}\\
  \end{array} \right).
\end{equation}
The continuously open quantum tribaker map 
is then given by means of the operator 
\begin{equation}\label{partialprojector}
 P=\left(\begin{array}{ccc}
  \openone_{N/3} & 0 & 0\\
  0 & \sqrt{F_R} \openone_{N/3} & 0\\
  0 & 0 & \openone_{N/3}\\
  \end{array} \right),
\end{equation}
 applied to Eq. (\ref{quantumbaker}) in such a way to preserve the original symmetries,  
\begin{equation}\label{partiallyopenquantumbaker}
 \widetilde{U^{\mathcal{B}}}= G_{N}^{-1} P G_{N/3} P.
\end{equation}

\begin{figure}
\includegraphics[width=8cm]{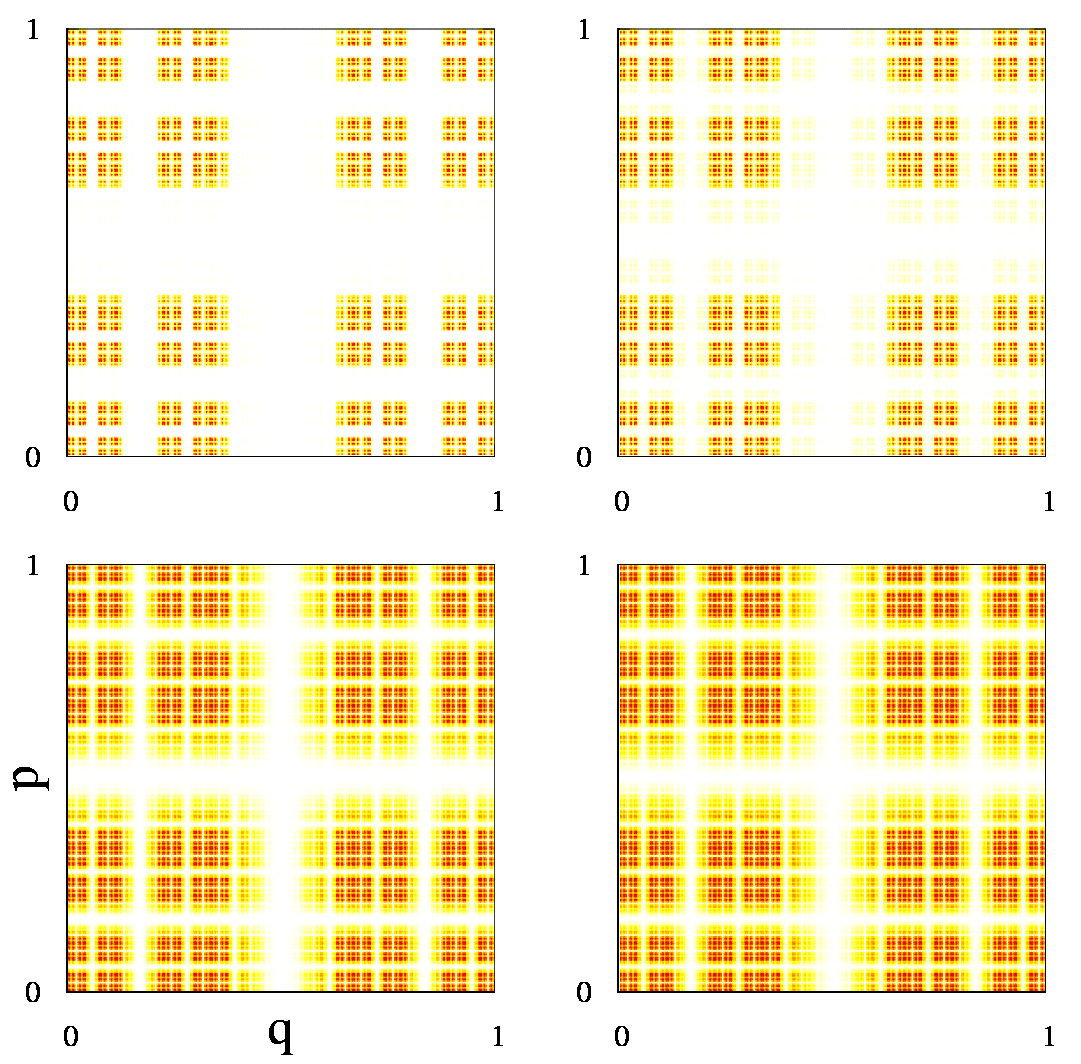}
\caption{(color online) Classical measure $\mu_{t,i}$ on the 2-torus for the 
continuously open tribaker map. In the upper panels we show the step opening, 
while in the lower ones the sinusoidal opening. In the left column $R=0.01$, and in the right 
one $R=0.1$.}
\label{fig1}
\end{figure}

Figure \ref{fig1} shows the finite time continuous repeller $\mu_{t,i}$ at time $t=10$. 
In the upper panels the step reflectivity is represented, while in the lower 
ones we find the sinusoidal case. In the left column $R=0.01$, and in the right one $R=0.1$.

\section{Results}
 \label{sec:results}

\begin{figure}
\includegraphics[width=8cm]{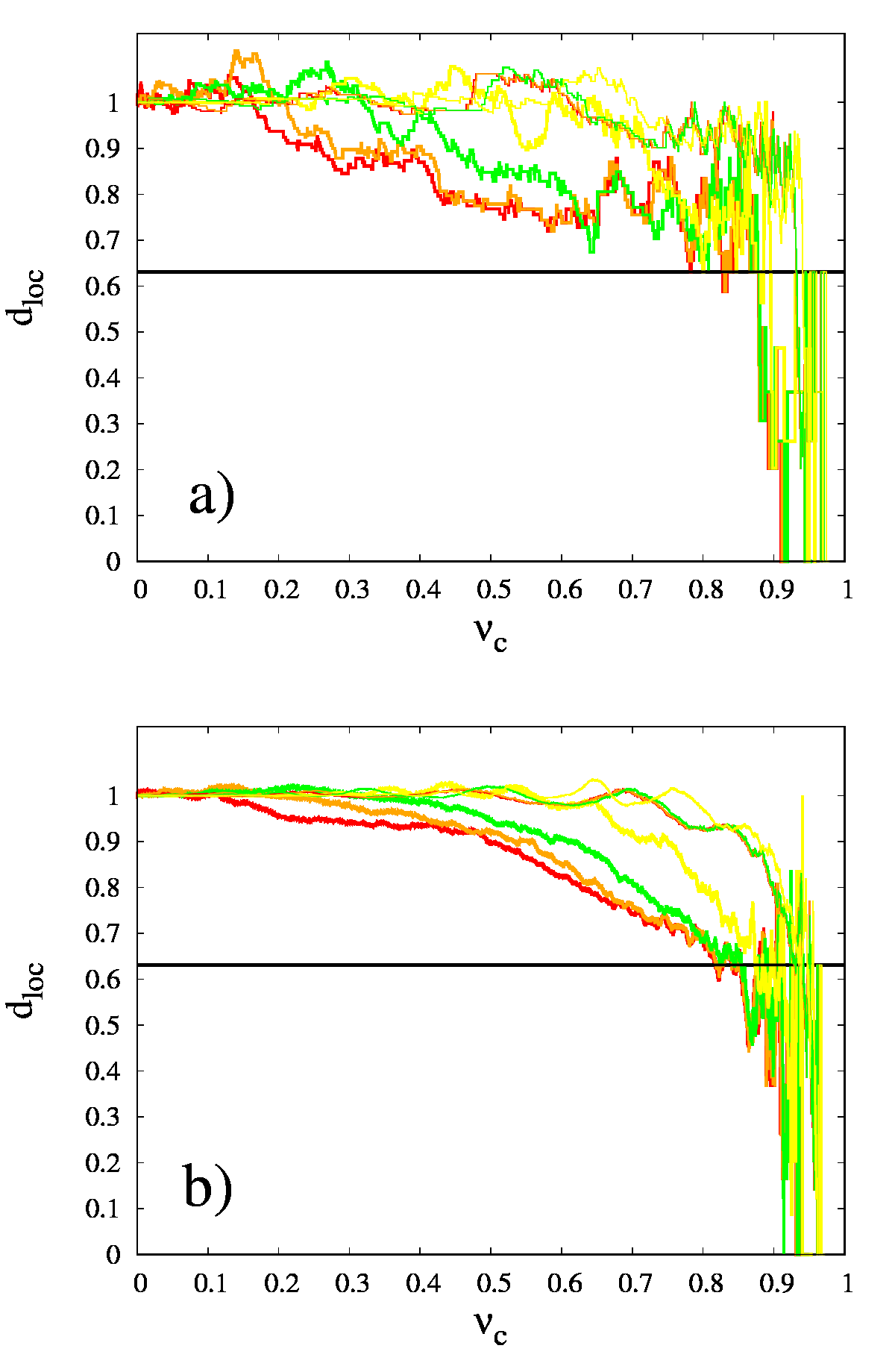}
\caption{(color online) Local dimension $d_{\rm loc}$ behaviour as a function of $\nu_{\rm c}$. Panel a) shows results 
for $N=3^5$, while panel b) corresponds to $N=3^9$. Thinner lines represent the sinusoidal opening and the 
thicker ones the step opening. By using darker to lighter shades of gray 
(red, orange, green and yellow) we display the cases with $R=0$, $R=0.001$, $R=0.01$, and $R=0.1$. The 
black horizontal line stands for the fractal dimension of the repeller (i.e. $\ln(2)/\ln(3)$).}
\label{fig2}
\end{figure}

 In order to characterize the spectral behaviour of continuously open maps we have evaluated 
 the local dimension $d_{\rm loc}=[\ln(M(N)) - \ln(M(N/3))]/\ln(3)$; $M(N)$ is the number of resonances 
 satisfying $\vert z_j \vert >\nu_{\rm c}$, and  we have chosen $N=3^5$ and $N=3^9$. This is a convenient quantity 
 that looks into the details of the spectral scaling behaviour \cite{Altmann}. We show the results 
 in Fig. \ref{fig2} for $R=0$, $R=0.001$, $R=0.01$, and $R=0.1$ (see caption for details).
It is clear that the marked oscillatory behaviour typical of the discontinuous opening \cite{Altmann} is 
almost completely absent in our cases. This seems to be valid even for the lowest $R$ values and in the large $N$ limit. 
This strong change in the spectral features suggests that the resonances follow 
a different kind of Weyl law (a different regime, at least). Nevertheless, there is a small portion 
of them above a given $\nu_{\rm c}$ that approximately follows a scaling ruled by the dimension 
of the repeller. For $N=3^5$, we take $\nu_{\rm c}=0.81$ for the step reflectivity case 
and $\nu_{\rm c}=0.91$ for the sinusoidal case.
 
\begin{figure}
\includegraphics[width=8cm]{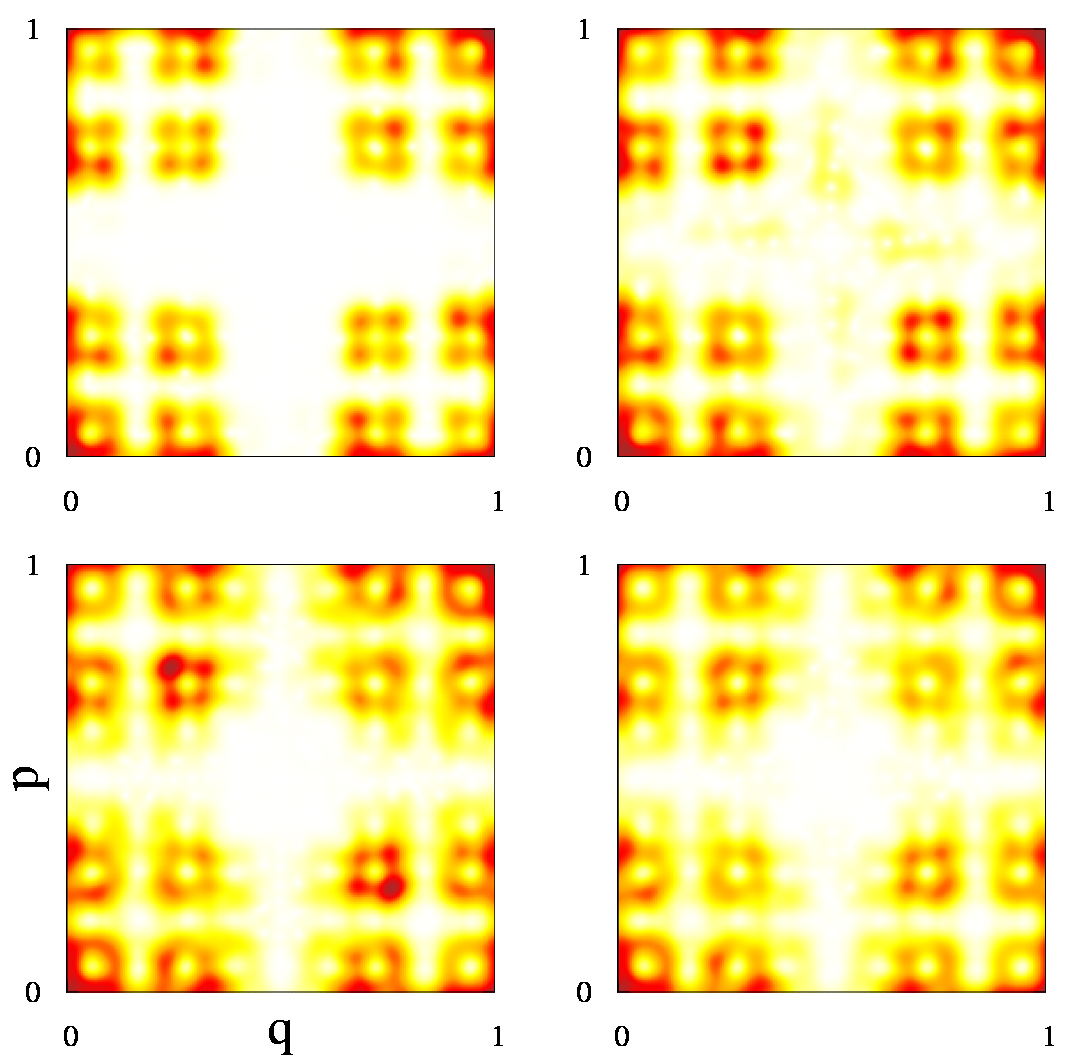}
\caption{(color online)  Semiclassical quantum continuous repeller $Q_{\nu_{\rm c}}^{sc}$ over the 
semiclassical long-lived resonances with eigenvalue moduli greater than $\nu_{\rm c}$. 
Upper panels correspond to the step opening ($\nu_{\rm c}=0.81$) 
and lower ones to the sinusoidal opening ($\nu_{\rm c}=0.91$). 
In the left column $R=0.01$, and in the right one $R=0.1$.}
\label{fig3}
\end{figure}

 Next, we apply the semiclassical theory to construct an approximation to the 
continuously open quantum tribaker map for $N=3^5$, several values of $R$, and 
considering POs up to period $L=7$.
A convenient quantum phase space representation can be obtained by means of the symmetrical 
operator $\hat{h}_j$ \cite{art0} associated 
to the right $|\Psi^R_j\ra$ and left $\la \Psi^L_j|$ long-lived semiclassical eigenstates, 
which are related to the eigenvalue $z_j$.  
\begin{equation}\label{eq.hdef}
 \hat{h}_j=\frac{\vert \Psi^R_j\rangle\langle \Psi^L_j\vert}{\langle \Psi^L_j\vert \Psi^R_j\rangle}.
\end{equation}
We sum the first $j$ of these projectors \cite{art3} (corresponding to the eigenvalues with 
the greatest moduli, $\vert z_j\vert\geqslant\vert z_{j^\prime}\vert$ with $j\leq j^\prime$) 
up to $\nu_{\rm c}$. 
\begin{equation}
\hat{Q}_j\equiv\sum_{j^\prime=1}^j\hat{h}_{j^\prime}.
\end{equation}
Their phase space representation by means of 
coherent states $\vert q,p\rangle$ is given by 
\begin{eqnarray}
 h_j(q,p)&=&\vert\langle q,p\vert \hat{h}_j\vert q,p\rangle\vert\\
 Q_j(q,p)&=& \vert\langle q,p\vert \hat{Q}_j\vert q,p\rangle\vert. 
\end{eqnarray}
This is the semiclassical quantum continuous repeller $Q_{\nu_{\rm c}}^{sc}$.
In Fig. \ref{fig3} we show the $Q_{\nu_{\rm c}}^{sc}$ obtained by using just the POs belonging 
to the repeller. The upper panels correspond to the step 
reflectivity function and the lower ones to the sinusoidal case. In the left column 
we have taken $R=0.01$, while in the right one $R=0.1$. 
The overlaps between these normalized distributions with the ones obtained by using the exact 
eigenstates, calculated as $O=\iint Q_{\nu_{\rm c}}(q,p) Q_{\nu_{\rm c}}^{sc}(q,p) dq dp$, 
are greater than $O=0.99$, in all cases.

Finally, and in order to examine the details of these quantum continuous repellers, 
we calculate the performance $P$ \cite{art3}, defined as the fraction of long-lived 
eigenvalues semiclassically reproduced within an error given by
$\epsilon=\sqrt{({\rm Re}{(z_i^{ex})}-{\rm Re}{(z_i^{sc})})^2+
({\rm Im}{(z_i^{ex})}-{\rm Im}{(z_i^{sc})})^2}$. In the latter expression 
$z_i^{ex}$ and $z_i^{sc}$ are the exact eigenvalues and those given 
by the semiclassical theory, respectively. 
We only consider the eigenvalues with modulus greater than $\nu_{\rm c}$.  
We calculate the number of scar functions $N_{SF}$ as a fraction of $N$ that are 
needed in order to obtain as many semiclassical eigenvalues inside the $\epsilon=0.001$ vicinity of the 
corresponding exact ones in order to reach $P \geq 0.8$. 
The fraction $N_{SF}/N$ is a measure of the morphology of the quantum continuous repeller. In fact, 
the larger this number the more interconnected the POs belonging to the open repeller are. 
In this sense, it quantifies the departure from the completely open case. 

\begin{figure}
\includegraphics[width=8cm]{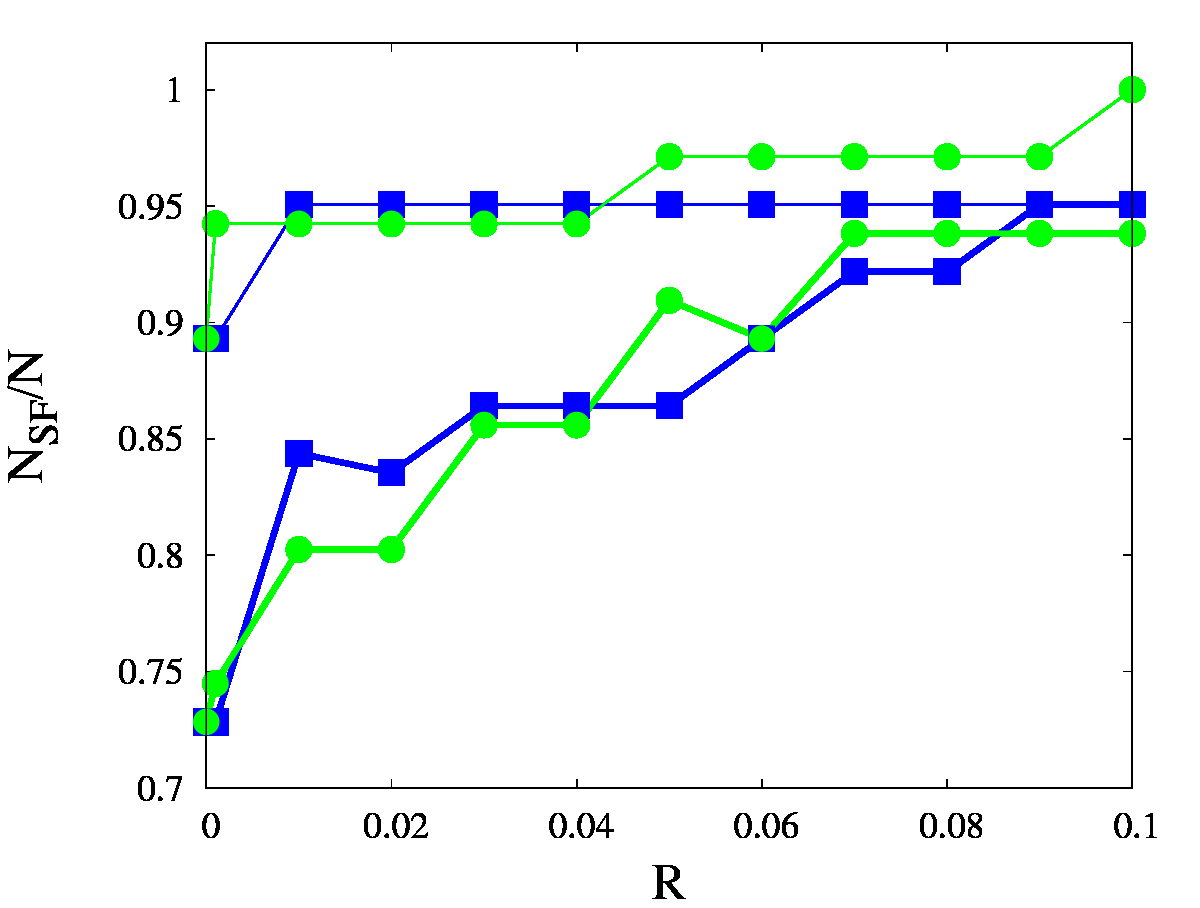}
\caption{(color online)  Fraction of scar functions $N_{SF}/N$ needed to reach 
$P=0.8$ as a function of the parameter $R$. The blue (black) lines with squares correspond 
to the cases considering only POs inside the repeller. Green (gray) lines with circles correspond 
to considering $N_{max}^{outPO}=5$ outside of it. Thinner lines represent the sinusoidal opening, 
while the thicker ones the step case.}
\label{fig4}
\end{figure}

In Fig. \ref{fig4} we show the fraction $N_{SF}/N$ needed to reach $P \geq 0.8$ 
as a function of $R \in [0:0.1]$. The upper and thinner lines correspond to the sinusoidal 
opening while the mainly lower and thicker ones to the step opening. The blue (black) lines with squares 
correspond to the case in which we only take POs that belong to the repeller. The green (gray) lines 
with circles display the results when taking into account a small maximum number of POs outside of 
the repeller, $N_{max}^{outPO}=5$. There is no significant improvement in the calculations when 
considering these POs, underlining the fact that the main role is played by the repeller.
It is also clear from Fig. \ref{fig4} that the step reflectivity function is able to keep the reduction 
in the number of POs needed that is typical in the discontinuous openings \cite{art4}. This is not the case 
for the sinusoidal function, though the spectral behaviour seems to be qualitatively similar in the sense 
that only small oscillations coming form the multifractal sampling are present \cite{Altmann}. 
It is interesting to notice that there is no important change that can be appreciated either in the classical 
and quantum continuous repellers or in the $d_{\rm loc}$ behaviour, between $R=0.01$ and $R=0.1$ in the sinusoidal 
opening; on the other hand the step opening shows clear differences.  

\section{Conclusions}
 \label{sec:conclusions}

 We have applied the recently developed short POs theory for partially open quantum maps to 
 the case where the reflectivity function is continuous on the phase space. This situation is 
 relevant for actual experiments with microresonators like those used to produce 
 microlasers \cite{Microlasers}. In particular, recent developments in this area show that 
 boundary conditions can be highly non trivial \cite{RecentExp}, and this of course impacts on many 
 properties. 
 
 By considering a Fermi-Dirac type step function we have tried to capture the essential features of continuity 
 vs. discontinuity, smoothing the sharp borders at the opening. On the other hand, we 
 have also used a sinusoidal function that models more generic situations. A parameter $R$ allows us to 
 control the degree of reflectivity by determining the bottom of these functions. 
 In both cases the spectral properties change with respect to the discontinuous opening. 
 The typical strong oscillations in the scaling of the number of resonances as a function of the eigenvalue 
 moduli above a given threshold $\nu_{\rm c}$ are almost completely lost, even for the lowest values 
 of the parameter $R$. This suggest that a different Weyl law (or regime) should be investigated for these cases. 
 
 Nevertheless and quite surprisingly, our semiclassical theory revealed that the role played by 
 the shortest POs that belong to the completely open repeller is very robust, at least 
 for $R<0.1$ and $N=3^5$, where we have made our calculations. These orbits give the fundamental 
 structure on which the quantum continuous repeller is constructed. This is underlined by the fact 
 that including a few POs outside of it does not significantly change the picture. Moreover, 
 in the case of the step opening we are able to use a relatively small amount of POs inside the 
 perturbative region. These findings could lead to new experiments in order to detect this structure, 
 and also to optimize the cavity design. In the future we plan to use this new insight on the morphology of the 
 eigenfunctions to find the scaling of the spectra and the reasons behind it.
 
\section{Acknowledgments}
The research leading to these results has received funding from CONICET (Argentina) under 
project PIP 112 201101 00703, and the  Ministerio de Econom\'\i a y Competitividad (MINECO) 
under contract MTM2015-63914-P, and by ICMAT Severo Ochoa SEV-2015-0554.

%
\end{document}